\documentclass[twocolumn,showpacs]{revtex4}
\usepackage{psfig}
\begin{document}
\title{
Okubo-Zweig-Iizuka rule violation in photoproduction}


\author{A. Sibirtsev$^{1}$, Ulf-G. Mei{\ss}ner$^{1,2}$, and A.W. Thomas$^3$}

\affiliation{
$^1$Helmholtz-Institut f\"ur Strahlen- und Kernphysik (Theorie), 
Universit\"at Bonn, Nu{\ss}allee 14-16, D-53115 Bonn, Germany \\
$^2$Institut f\"ur Kernphysik (Theorie), Forschungszentrum J\"ulich,
D-52425 J\"ulich, Germany,\\
$^3$Jefferson Lab, 12000 Jefferson Ave., Newport News, VA 23606, USA
}

\begin{abstract}
We investigate OZI rule violation in $\omega$ and $\phi$-meson
photoproduction off nucleons. Data on the total cross sections indicate a large
$\phi{/}\omega$ ratio of about 0.8 at the maximal available
photon energy that is in good agreement with expectations from QCD.
On the other hand, data at large four momentum transfer 
exhibit a ratio of about 0.07, showing 
that the perturbative QCD regime is not approached at 
$|t|{>}$2~GeV$^2$ and photon energies $E_\gamma{<}$4~GeV. 
The anomanously large $\phi{/}\omega$ ratio at low energies, 
that is close to the reaction threshold, remains to be explained within
nonperturbative QCD. 
\vspace{-6.5cm}

\hfill {\tiny FZJ--IKP(TH)--2005--10, 
HISKP-TH-05-07, JLAB-THY-05-309}

\vspace{6.5cm}
\end{abstract}

\maketitle

The violation of the OZI rule has been one of the more challenging aspects of
QCD since the famous conjecture of Okubo, Zweig and 
Iizuka~\cite{Okubo,Zweig,Iizuka} based on the breaking of SU(3)
symmetry. The $\phi$-meson can be considered as a pure $s{\bar s}$ 
quark state if the octet and singlet mesons are ideally mixed with the
angle $\theta_V$=35.3$^\circ$. Real life is
not ideal and the experimental deviation from the ideal mixing angle
is $\Delta\theta_V$=3.7$^\circ$~\cite{PDG}. As a result, the 
$\phi$-meson contains light 
quarks and the ratio of $\phi$ to $\omega$-meson production in different 
reactions containing strange as well as non-strange quarks, 
such as $K^- p  \to VY$, $\pi N \to VX$ or $NN \to V X$ ($V=
\omega,\phi)$,
can be estimated~\cite{Lipkin} as 
R$_{\phi{/}\omega}{=}\tan^2\Delta\theta_V{\simeq}4.2{\times}10^{-3}$. 
The deviation of 
R$_{\phi{/}\omega}$ from zero in such reactions is usually refered 
to  as OZI rule violation. One might expect an even larger ${\phi{/}\omega}$  
ratio from  reactions involving nucleons 
because of the intrinsic $s{\bar s}$ 
content of the nucleon. In that case the strangeness component
of the initial  nucleon can be transfered to the final $\phi$-meson.
For a nice review on the OZI rule and its experimental tests, see
\cite{Nomokonov:2002jb}.

A systematic analysis~\cite{Sibirtsev1} of available data
on $\phi$ and $\omega$-meson production in $pp$ and $\pi{p}$ reactions
gives R$_{\phi{/}\omega}{\simeq}(13.4{\pm}3.2){\times}10^{-3}$.
This large ratio was interpreted~\cite{Sibirtsev1} in terms  
$pp$ and $\pi{p}$ reaction dynamics that involves the $\phi\rho\pi$ 
and $\omega\rho\pi$ vertices. The $\phi\rho\pi$ 
coupling constant can be  evaluated directly from the $\phi{\to}\rho\pi$ 
decay. The  $\omega\rho\pi$ coupling can be extracted from
$\omega{\to}3\pi$ decay~\cite{Gellmann} or from $\omega{\to}\pi\gamma$
and $\rho{\to}\pi\gamma$ decays~\cite{Meissner1}, that are dominated
by the $\omega{\to}\rho\pi$ vertex with an intermediate vector meson coupled
to the photon. The different decay modes provide~\cite{Sibirtsev1} 
an average ratio of
R$_{\phi{/}\omega}{\simeq}(12.5{\pm}3.4){\times}10^{-3}$, which is 
in a good agreement with available results from $pp$ and $\pi{p}$ reactions.
It is clear that the large experimental ${\phi{/}\omega}$ 
ratio is dictated by the large $\phi{\to}\rho\pi$ decay width 
and is not related to the strangeness content of the nucleon. 
It is well known~\cite{Jain} that the $\phi{\to}\rho\pi$ decay violates
the OZI rule. We remark  that new experiments
with the ANKE detector at COSY~\cite{Hartmann} and at the 
JINR Nuclotron~\cite{Salmin} are  devoted to the investigation of    
OZI rule violation at energies close to the $pp{\to}pp\phi$ 
reaction threshold. Here,  we present the current status of OZI rule
violation in vector meson photoproduction from nucleons.

\begin{figure}[t]
\vspace*{-7mm}
\centerline{\hspace*{2mm}\psfig{file=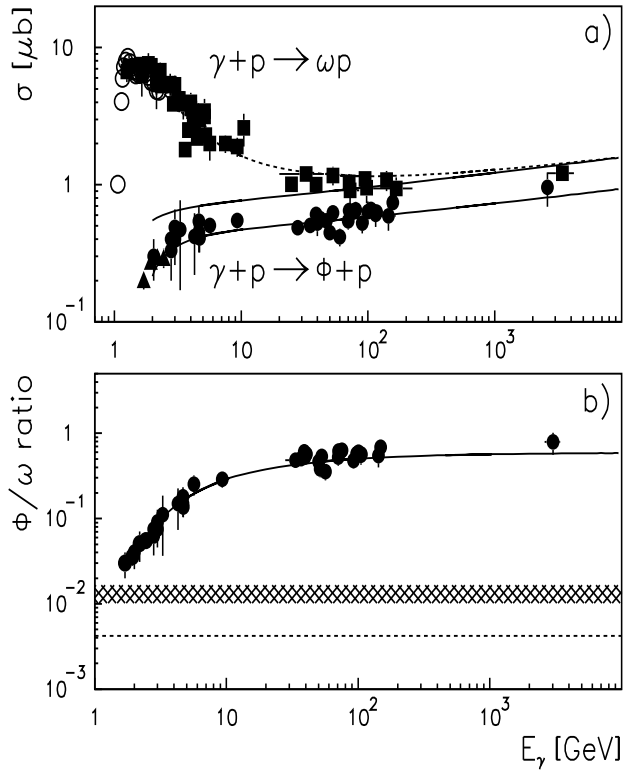,width=9.4cm,height=11.5cm}}
\vspace*{-5mm}
\caption{a) Total cross section for exclusive $\omega$ and $\phi$
photoproduction off the proton as a function of the photon 
energy. The solid squares and circles show the data collected
in Refs.~\cite{Busenitz,Derrick,Landolt,Sibirtsev3}. 
The triangles and open circles show most recent 
results from ELSA~\cite{Barth1,Barth2} for $\phi$ and $\omega$-meson 
photoproduction respectively.  The solid lines give
the result from pomeron exchange, while the dashed line
shows the calculations for the $\omega$-meson with both meson and 
pomeron exchange contributions. b) The ratio R$_{\phi{/}\omega}$.
The dashed line is the ratio given by SU(3) mixing, the shaded area
indicates results from $\pi{p}$ and $pp$ reactions and the solid line
is the ratio of the calculations presented in a).}
\label{rosi4}
\end{figure}

The solid squares and circles in Fig.\ref{rosi4}a) show the 
available experimental results on the total cross 
section for $\omega$ and $\phi$-meson photoproduction off a proton 
as a function of photon energy~\cite{Busenitz,Derrick,Landolt,Sibirtsev3}.
The triangles in Fig.\ref{rosi4}a) are the most recent results on
$\phi$-meson photoproduction at $E_\gamma{<}$2.6~GeV measured 
at ELSA~\cite{Barth1}, while the open circles show ELSA
results for $\omega$-photoproduction~\cite{Barth2}.
The solid lines indicate  calculations based on  the 
pomeron exchange model~\cite{Donnachie1,Sibirtsev2},
\begin{eqnarray}
\frac{d\sigma}{dt} &=& \frac{81  m_V^2 \beta_0^4 \mu_0^4 \,\Gamma}
{\pi\alpha}\left[\frac{s}{s_0}\right]^{2\alpha_P(t)-2}\nonumber\\
&\times& \frac{F^2(t)}{(m_V^2-t)^2(2\mu_0^2+m_V^2-t)^2},
\label{pom}
\end{eqnarray}
where $t$ and $s$ are the four-momentum transfer and the invariant
collision energy squared, respectively, $m_V$ is the mass of the vector
meson $V$, $\Gamma$ is the $V{\to}e^+e^-$ decay width, $\alpha$ is 
the fine-structure constant and $F(t)$ is the proton isoscalar 
electromagnetic form factor, approximated as
\begin{eqnarray}
F(t)=\frac{4m_p^2-2.8t}{4m_p^2-t} \frac{1}{(1-t/t_0)^2}~.
\end{eqnarray}
Here,  $m_p$ is the proton mass and $t_0$=0.7~GeV$^2$. The free 
parameters of the model are related to  the pomeron-quark
vertex, namely the coupling constant, $\beta_0$, and the cut-off of the
form factor, $\mu_0$. The pomeron trajectory $\alpha_P(t)$ is
given by~\cite{Donnachie2}
\begin{eqnarray}
\alpha_P(t)=\alpha_0+\alpha^\prime t\, ,
\end{eqnarray}
where $\alpha_0$=1.008 and $\alpha^\prime$=0.25~GeV$^{-2}$.
The constant $s_0{=}1{/}\alpha^\prime$ was determined utilizing the
dual model prescription~\cite{Veneziano}. 
The parameter $\mu_0^2$=1.1~GeV was 
evaluated~\cite{Donnachie1,Donnachie2} from 
high energy data on elastic and inelastic scatterings at small 
$|t|$. A systematic analysis~\cite{Sibirtsev3} of total and differential 
cross sections for $\omega$-meson photoproduction results in 
$\beta_0$=2.35~GeV$^{-1}$. Fig.\ref{rosi4}a) illustrates that 
pomeron exchange describes  the data on $\omega$-meson 
photoproduction at photon energies above  100~GeV quite well. To reproduce 
the data at lower energies it is necessary to account~\cite{Sibirtsev3} 
for both  meson exchange and pomeron contributions -- as 
shown by the dashed line in Fig.\ref{rosi4}a).
To reproduce the total cross section for $\phi$-meson photoproduction
we readjust  $\beta_0$=1.9~GeV$^{-1}$. The pomeron exchange alone
describes the total cross section of the $\gamma{p}{\to}\phi{p}$ 
reaction even at low energies.

In Fig.~\ref{rosi4}b) we display the data on the ratio of $\phi$ to $\omega$ 
total photoproduction cross sections together with the calculations
presented in Fig.\ref{rosi4}a). The dashed line in Fig.\ref{rosi4}b)
is the ratio R$_{\phi{/}\omega}{=}4.2{\times}10^{-3}$, given by the 
octet and singlet mixing, while the shaded strip shows the result
from $\pi{p}$ and $pp$ reactions. Apparently the OZI rule is strongly
violated in photoproduction. At high energies the interaction is
driven by gluon exchange and is flavor-blind. Thus we would expect that
in the perturbative QCD regime the ${\phi{/}\omega}$ ratio might 
approach unity up 
to the corrections associated with the hadronic 
wave functions of the $\omega$ and
$\phi$-mesons. Indeed the data collected in  Fig.\ref{rosi4}b)
show that  R$_{\phi{/}\omega}{\simeq}0.5$ and that it does not depend on 
the photon energy at 30 ${\leq}E_\gamma{<}$200~GeV. At the maximal 
available photon
energy R$_{\phi{/}\omega}{\simeq}0.8{\pm}0.2$.
This result is in  reasonable agreement with the non-perturbative
QCD quark-pomeron interaction that should  also be flavor-blind.
Although we found that in reality the quark-pomeron coupling
depends on the quark flavor and $\beta_0$=2.35, 1.9 and 
0.45~GeV$^{-1}$ for light, strange and charm~\cite{Sibirtsev2} quarks, 
respectively, the data on the large  ${\phi{/}\omega}$ ratio  
support the dominance of QCD interactions at high energies.

Furthermore, the photoproduction data indicate a large ratio 
R$_{\phi{/}\omega}{>}$0.03 even at threshold. 
The agreement between the $\phi$-meson photoproduction data and pomeron 
calculations at low energies does not provide a reasonable 
explanation of the large ratio.  It is known that the pomeron theory is 
applicable~\cite{Donnachie2} in the high energy region, 
$E_\gamma{>}$10~GeV, and an agreement between the data and 
calculations at low energies might be rather accidental. 
For instance, a systematic
analysis~\cite{Sibirtsev2,Sibirtsev4} of $ J/\Psi$-photoproduction 
also indicates good agreement between the total photoproduction 
cross section and pomeron exchange at threshold, but at the same 
time shows a strong disagreement between the calculated and the 
measured $ J/\Psi$-meson differential spectra. Apparently a systematic
analysis of $\phi$-meson photoproduction is necessary to 
understand the reaction mechanism  and the
anomalously large ratio R$_{\phi{/}\omega}$  at low energies.

It is important that the QCD regime be studied at high 
energies and at large four momentum
transfer squared. At large $|t|$ the interaction probes small
distances ${\simeq}1/\sqrt{-t}$ and can be decribed by multi--gluon 
exchange and constituent quark interchange~\cite{Gunion,Landshoff,Brodsky1}.
Because of the OZI rule, quark interchange does not contribute
to $\phi$-meson photoproduction and one might expect that the
$\gamma{p}{\to}\phi{p}$ reaction at large $|t|$ would be dominated by 
gluon exchange.  Indeed, the two-gluon exchange 
model~\cite{Donnachie3,Laget} reproduces data on $\phi$-meson 
photoproduction at large $|t|$. On the contrary,  $\omega$-photoproduction
allows for  quark interchange. This might result in a small value for  
R$_{\phi{/}\omega}$. If both
$\omega$ and $\phi$-meson photoproduction are dominated by gluon exchange,
we might expect a large R$_{\phi{/}\omega}$, compatible 
with the result shown in Fig.\ref{rosi4}b) at high photon energies. It is 
worthwhile to note the analogy between the phenomenological pomeron
and two-gluon exchange~\cite{Donnachie1,Cudell}, which allows for 
explicit comparison of large $|t|$ and high energy data.

\begin{figure}[t]
\vspace*{-7mm}
\centerline{\hspace*{2mm}\psfig{file=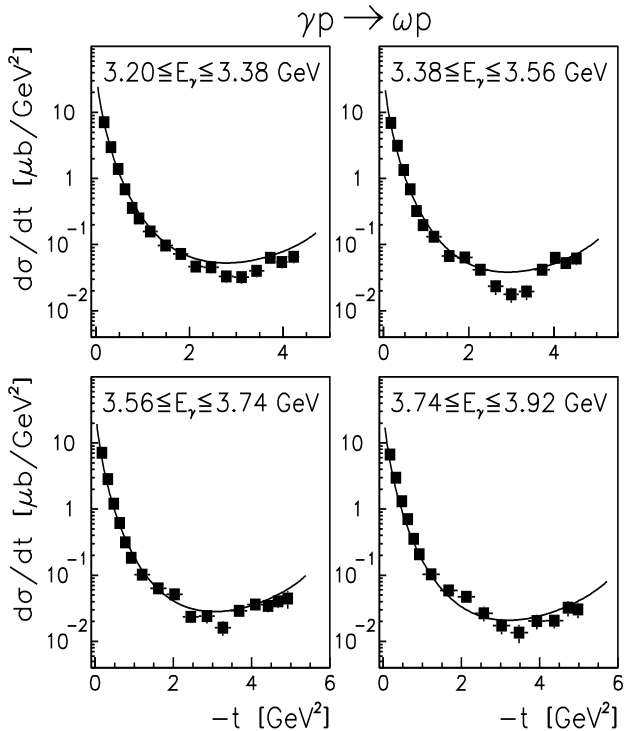,width=9.4cm,height=10.6cm}}
\vspace*{-5mm}
\caption{Differential cross section for $\gamma{p}{\to}\omega{p}$
reaction as a function of four momentum transfer squared~\cite{Battaglieri}.
The lines show the meson and nucleon exchange 
calculations~\cite{Sibirtsev5}.}
\label{rosi2}
\end{figure}

\begin{figure}[t]
\vspace*{-12mm}
\centerline{\hspace*{2mm}\psfig{file=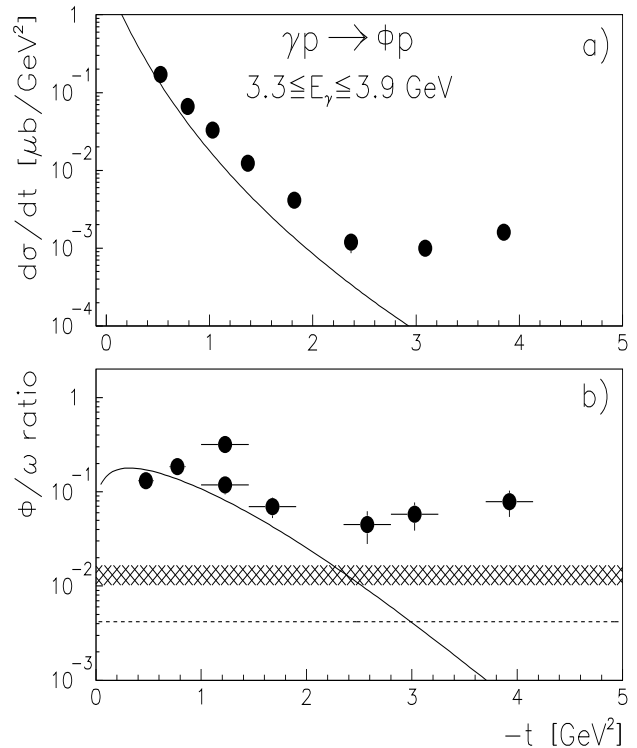,width=9.4cm,height=11.5cm}}
\vspace*{-7mm}
\caption{a) Differential cross section for the $\gamma{p}{\to}\phi{p}$
reaction as a function of four-momentum transfer squared~\cite{Anciant}.
The solid lines represent the calculations based on pomeron exchange. b) The 
$\phi{/}\omega$ ratio as function of $-t$. The dashed line is the 
ratio given by SU(3) mixing, the shaded area
indicates the results from $\pi{p}$ and $pp$ reactions and the solid line
is the ratio of the calculations with meson and nucleon exchanges  
in the $\omega$ case and with pomeron exchange for $\phi$-photoproduction.}
\label{rosi3}
\end{figure}

High accuracy data on $\omega$ and $\phi$-meson photoproduction at
large $|t|$ were collected by the CLAS Collaboration at 
JLab~\cite{Anciant,Battaglieri} and are shown in 
Fig.\ref{rosi2} and Fig.\ref{rosi3}a). The solid lines in 
Fig.\ref{rosi2} show the calculation based on meson and nucleon exchange, 
with vertex parameters fixed by data available prior to the CLAS
measurements. Here the $\omega$-meson photoproduction at large $|t|$
is dominated by the nucleon exchange current, while at low $|t|$
the dominant contribution comes from $\pi$ and $\sigma$-exchange.
The solid line in Fig.\ref{rosi3}a) represents the calculations based
on  pomeron exchange utilizing Eq.(\ref{pom}). 
Fig.\ref{rosi3}b) shows the ${\phi{/}\omega}$ ratio
as a function of $|t|$. 

Let us first discuss the result at low $|t|$ where 
R$_{\phi{/}\omega}{\simeq}$0.15.  As we mentioned before, 
the application of the pomeron exchange at low energies can
not be justified theoretically~\cite{Donnachie2,Sibirtsev2,Sibirtsev3}
and one might consider the contribution from $\pi$ and 
$\sigma$-exchange, as in $\omega$-meson photoproduction.
In that case the ${\phi{/}\omega}$ ratio depends on the ratio
of the $\phi\pi\gamma$ and  $\omega\pi\gamma$ coupling constants 
squared, which is $\simeq(3.5{\pm}0.2)10^{-3}$~\cite{Meissner1,Jain}, 
close to the ratio given by SU(3) mixing. We would not expect 
$\sigma$-exchange to contribute much more to $\phi$-meson
photoproduction in comparison to $\omega$-photoproduction. At least 
this is not supported by estimates given by $\phi{\to}\pi\pi\gamma$ and
$\omega{\to}\pi\pi\gamma$ decays~\cite{PDG}. The $\eta$-exchange also
plays a minor role~\cite{Sibirtsev5}.

\begin{figure}[t]
\vspace*{-6mm}
\centerline{\hspace*{2mm}\psfig{file=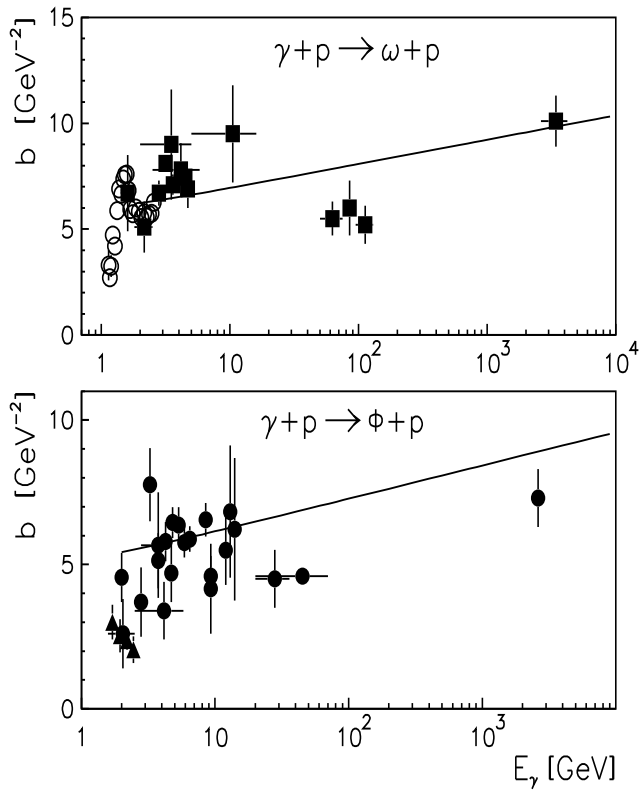,width=9.6cm,height=11.5cm}}
\vspace*{-6mm}
\caption{Exponential slope of the $t$-dependence for exclusive
$\omega$ and $\phi$-photoproduction as a function of the photon 
energy. Solid squares and circles show the data collected in 
Refs.~\cite{Derrick,Sibirtsev3,Erbe,Ballam,Behrend,Aston,Atkinson}.
The open circles and triangles indicate recent results 
measured at ELSA~\cite{Barth1,Barth2} for $\omega$ and 
$\phi$-meson photoproduction, respectively.
Solid lines show the result of the pomeron exchange model.}
\label{rosi5}
\end{figure}

Therefore either one would 
accept that pomeron exchange already dominates $\phi$-photoproduction  
at threshold or we must consider another possible nonperturbative 
mechanism. For instance near threshold $J/\Psi$-meson photoproduction
was discussed in terms of three gluons~\cite{Brodsky2} or 
anomalous axial exchange~\cite{Sibirtsev4}. While neither  mechanism
depends strongly on the photon energy, they have a very different low $|t|$
dependence, which can be used for an experimental falsification of 
such models. For illustration we show in Fig.~\ref{rosi5} the exponential 
slope of the $t$-dependence for exclusive $\omega$ and $\phi$-meson 
photoproduction. The solid lines indicate the slope fitted to pomeron 
exchange calculations at $|t|{<}$0.6~GeV$^2$. At low energies the slope of
$\gamma{p}{\to}\phi{p}$ data differs from the pomeron calculation
where the minimal slope is driven by the proton isoscalar electromagnetic
form factor squared~\cite{Sibirtsev2}. The data indicate a soft
contribution with a slope around 4~GeV$^{-2}$ at energies
$E_\gamma{<}$10~GeV. The open circles and triangles in Fig.~\ref{rosi5}
shows most recent high precision results from ELSA~\cite{Barth1,Barth2}
for $\omega$ and $\phi$-photoproduction, respectively.  The ELSA
data  apparently indicate a small slope and are inconsistent with
pomeron exchange. Moreover the ELSA $\phi$-meson photoproduction
measurements~\cite{Barth1} of the spin density matrix elements  
clearly contradicts the expectations based on the pomeron model.
However, apart from the ELSA measurements, the data at $E_\gamma{<}$10~GeV
are not sufficiently accurate 
to draw more solid conclusions. In that respect, new precise differential 
cross section data  from ELSA, JLab and SPRING8 on $\phi$-meson 
photoproduction  are of  great importance. 

Furthermore the $\phi{/}\omega$ ratio at large $|t|$, 
R$_{\phi{/}\omega}{\simeq}0.07{\pm}0.01$, differs substantially
from that at high energies, which is in disagreement with
the two-gluon exchange model. Other nonperturbative mechanisms, such as 
(for example) multi-gluon exchange are also not supported by the data, 
since any 
flavour-blind interaction would result in a large ratio close to the
R$_{\phi{/}\omega}{\simeq}0.5$ observed at high energies in the
QCD regime. 

Alternatively one might consider nucleon exchange, which has been shown 
to contribute  
to $\omega$-photoproduction at large-$|t|$, as illustrated by
Fig.~\ref{rosi2} and might also contribute to $\phi$-meson
photoproduction. In that case the $\phi{/}\omega$ ratio 
at the same $t$ would be given explicitly by the ratio of 
the $\phi{pp}$ and $\omega{pp}$ coupling constants 
squared. Modern dispersion theoretical 
analysis~\cite{Hammer,Mergell} of the nucleon electromagnetic form factors 
shows that the squared ratio of the $\phi{pp}$ to $\omega{pp}$ 
vector couplings is ${\simeq}0.23$. The dispersion analysis 
is based on a maximal violation of the OZI rule because the isoscalar
spectral function in the mass region above 1~GeV was taken to come 
entirely from the $\phi$-pole. The combined analysis ~\cite{Meissner2}
of the results from the nucleon-nucleon interaction and 
nucleon electromagnetic 
form factors led to a  substantially smaller ratio, of order  
$5.3{\times}10^{-4}$. The analysis~\cite{Haidenbauer,Nakayama} of 
the angular $\phi$-meson spectra from the $pp$ interaction provides
an upper limit of the squared ratio as ${\simeq}0.024$.  
Obviously within these large uncertainties of the ratio 
for the $\phi{pp}$ and $\omega{pp}$ couplings one might 
describe $\phi$-meson photoproduction at large $|t|$ by  
nucleon exchange.

To summarize, we have given a systematic analysis of OZI rule violation
in vector meson photoproduction. The data on the total cross sections 
for $\phi$ and $\omega$-meson photoproduction indicate a 
ratio R$_{\phi{/}\omega}{\simeq}0.8{\pm}0.2$ at the maximal available
photon energy. This large ratio is in agreement with pomeron or
two-gluon exchange calculations and fulfills  QCD
expectations  at high energies. Moreover, we found that the $\phi{/}\omega$
ratio is already greater than 0.03 at threshold -- that is, it 
substantially exceeds the ratio  R$_{\phi{/}\omega}{=}4.2{\times}10^{-3}$ 
expected from SU(3) mixing. At low photon energies this anomalously large
ratio can be explained if one assumes that the $\phi$-meson production
is already entirely dominated by pomeron exchange at threshold.

On the other hand, the available data at large four-momentum transfer squared
indicate that the QCD regime is not approached at $|t|{>}$2~GeV$^2$
and $E_\gamma{<}$4~GeV. Here the ratio 
R$_{\phi{/}\omega}{\simeq}0.07{\pm}0.01$ differs from the results
observed at high energies. We speculate that at low photon energies and
large $|t|$ the $\phi{/}\omega$ ratio might be explained by
the nucleon exchange mechanism. At low $|t|$ the ratio approaches
$\simeq$0.15 and can once again be understood in terms of the dominance of
pomeron or two-gluon exchange in $\phi$-meson photoproduction
at low photon energies. However, such a dominance is in
contradiction with results available for the exponential slope of the
$t$-dependence, which indicate a soft component at low-$|t|$
photoproduction at $E_\gamma{<}$10~GeV. The soft component  rules
out the pomeron exchange model at low energies because
of  the coupling of the pomeron  to the
isoscalar electromagnetic form factor of the proton, which
results in a large slope~\cite{Sibirtsev2}.

In conclusion, while all available data for $\omega$ and $\phi$-meson 
photoproduction explicitly indicate a substantial violation of the OZI 
rule and show a $\phi{/}\omega$ ratio much larger than that 
obtained~\cite{Sibirtsev1} from $\pi{p}$ and $pp$ reactions,
only part of our findings can be explained in terms of perturbative 
QCD. The large $\phi{/}\omega$ ratio at low energies remains a 
puzzle of nonperturbative QCD, see also \cite{KV}.
Further progress apparently requires
new, precise data on $\phi$-meson photoproduction close to the
reaction threshold, as well as intensive theoretical activity
to develop new QCD methods at low energies.

\vspace{-5mm}

\acknowledgments{

\vspace{-2mm}

We appreciate discussions  with
M.~Battaglieri, J.~Haidenbauer, C.~Hanhart, M.~Hartmann,
F.~Klein and S.~Krewald. 
This work was partially supported by the Department of
Energy under contract DE-AC05-84ER40150 under which SURA operates
Jefferson Lab, by Deutsche Forschungsgemeinschaft through funds
provided to the SFB/TR 16 ``Subnuclear Structure of Matter'' and
by the COSY FFE grant No. 41445400 (COSY-067). 
This research is part of the EU Integrated Infrastructure Initiative Hadron
Physics Project under contract number RII3-CT-2004-506078.
}

\end{document}